\documentclass[a4paper,12pt]{article}

\newcommand{\RR}{{\rm R}}
\newcommand{\RRR}{\RR^{1,3}}
\newcommand{\RRRR}{\RR^{4}}
\newcommand{\CC}{{\cal C}}
\newcommand{\half}{{\textstyle{1\over2}}}

\newcommand{\utez}{w}
\newcommand{\prep}{u}
\newcommand{\retfakt}{s}

\newcommand{\Polje}{\Phi}
\newcommand{\polje}{\phi}
\newcommand{\polj}{\varphi}
\newcommand{\poljF}{\widetilde{\polj}}

\newcommand{\polF}{\polje_F}

\newcommand{\PolX}{\Polje}

\newcommand{\PolS}{\Polje_A}
\newcommand{\PolSe}{\Polje_{A1}}

\newcommand{\polSe}{\polj_{A1}}
\newcommand{\polSFe}{\widetilde{\polj}_{A1}}

\newcommand{\poljeR}{\polje_{ret}}
\newcommand{\polR}{\polj_{ret}}

\newcommand{\PolR}{\Polje_{ret}}

\newcommand{\akcija}{I}

\newcommand{\akcijaA}{\akcija_A}

\newcommand{\izvor}{J}
\newcommand{\izvorF}{\widetilde{J}}

\newcommand{\Lag}{{\cal L}}
\newcommand{\Lagf}{\Lag_{0}}
\newcommand{\Lagi}{V}

\newcommand{\LagAf}{\Lag_{A0}}

\newcommand{\fakt}{f}

\newcommand{\propO}{\Delta}
\newcommand{\propA}{\Delta_A}
\newcommand{\propAe}{\Delta_{A1}}
\newcommand{\propQ}{\Delta_F}

\newcommand{\propF}{\widetilde{\Delta}}
\newcommand{\propFF}{\propF_F}
\newcommand{\propAF}{\propF_A}
\newcommand{\propAFe}{\propF_{A1}}

\newcommand{\Gf}{G^{(n)}}

\newcommand{\GfA}{\Gf_A}
\newcommand{\GFAT}{G_A^{(2)}}

\newcommand{\SmA}{S_A}

\title{Two basic problems posed by quantum scattering of fundamental particles}
\author{Marijan Ribari\v c and Luka \v Su\v ster\v si\v c\thanks{Corresponding author. Phone +386 1 477 3258; fax +386 1 423 1569; electronic address: \tt luka.sustersic@ijs.si\rm} \\Jo\v zef Stefan Institute, p.p.3000, 1001 Ljubljana, Slovenia }
\date{}

\begin{document}

\maketitle

\begin{abstract}
We put forward a finite theory of quantum scattering of fundamental particles without using auxiliary particles. It suggests that to avoid ultraviolet divergencies and model faster-than-light effects it suffices to appropriately change only the free-field Lagrangians while retaining their locality in space-time and Lorentz invariance. Using functions of two independent four-vector variables, we base this finite theory on the path-integral formalism on the four-dimensional space-time and the Lehmann-Symanzik-Zimmermann reduction formula.
\end{abstract}
\vfill
\vskip0.1cm\noindent UDC: 530.145, 530.12
\vfill\eject

\section{Introduction}
\label{uvod}

In modeling the quantum scattering of fundamental particles (quantum scattering for short), there are two more than sixty years old problems concerning \it conceptual consistency \rm and \it physical completeness \rm of the model: 
\begin{itemize}
\item[(A)]If we calculate the results of quantum scattering by a quantum field theory (QFT), perturbative expansions of $n$-point Green functions result, in general, in ultravioletly divergent integrals. Formal regularization methods one resorts to to make these integrals finite are not conceptually consistent with the established concepts of theoretical physics, see e.g.\ \cite{Dirac} and \cite{Weinberg} subsection~1.3. 
\item[(B)]If we go beyond a strictly operational interpretation of quantum scattering, certain results suggest the existence of arbitrary fast\footnote{We will refer to an effect (i.e. a change in the state of a physical system) that occurs at the space-time point $(ct_2, \vec{r}_2)$ owing to a source at $(ct_1, \vec{r}_1)$ as: \it retarded, \rm if $t_2 > t_1$; \it relativistic, \rm if it is retarded and the Einstein causality condition $| \vec{r}_2 - \vec{r}_1 | / (t_2 - t_1) \le c$ is satisfied; \it arbitrary fast, \rm if it is retarded and $| \vec{r}_2 - \vec{r}_1| / (t_2 - t_1) > c$ could be arbitrary large; and \it backward-in-time, \rm if $t_2 < t_1$.} effects (AFEs for short).\footnote{Heisenberg noted already in 1930 that this kind of effects can never be utilized to transmit signals so that their existence could not be in conflict with the Einstein relativity postulates, cf.~\cite{Eberh1} \ subsection~2.1.1.} This was brought up already by the Einstein, Podolski and Rosen (EPR) thought experiment, and was experimentally verified by violations of the Bell inequalities. As far as we know, there is no model of quantum scattering and associated AFEs consistent with Einstein's relativity postulates\footnote{As Kacser pointed out \cite{Kacser}, one can base special relativity on ``Einstein's \it first relativity postulate: \rm The basic laws of physics are identical for two observers who have a constant relative velocity with respect to each other'' (i.e. there is no fundamental frame of reference) and on ``the \it second relativity postulate: \rm the speed of light in vacuum is an absolute constant for all observers, independent of the velocity of the light source, or the velocity of the observer''.}. For some comments on faster-than-light effects, as AFEs are usually refered to, see e.g.~\cite{Bell, Eberh1, semanj, nevem}, and references therein. 
\end{itemize}
To avoid non-essential calculational complications, we will consider these two problems in the case of modelling the quantum scattering of identical spin 0 particles by the trivial QFT\footnote{According to Brunetti and Fredenhagen \cite{Brunetti}, ``The quest of the existence of a non-trivial quantum field theory in four space-time dimensions is still without any conclusive result''.} of a single real scalar field $\polje(x)$ with $\polje^4$ self-interaction. As the QFT action functional we take
\begin{equation}
   \akcija[\polje; J] = \int \! d^4x \,\Bigl ( \Lagf(\polje, \partial_\mu \polje) - \Lagi(\polje) + \polje J \Bigr) \,,
   \label{QFTaction}
\end{equation}
where the QFT free-field Lagrangian (density) 
\begin{equation}
   \Lagf(\polje, \partial_\mu \polje) =   - \half (\partial_\mu \polje)^2 - \half m^2 \polje^2  \,;
   \label{QFTfreelagr}
\end{equation}
the interaction Lagrangian
\begin{equation}
   \Lagi(\polje ) = {\lambda \over 4!} \polje^4 + {Z^2\lambda_0 -\lambda \over 4!} \polje^4 
               + {Z - 1\over 2} (\partial_\mu \polje)^2 + {Z m_0^2 - m^2\over 2} \polje^2 \,;
   \label{QFTintlagr}
\end{equation}
$m$ and $\lambda$ are real non-negative coefficients; $m_0$, $Z$, and $\lambda_0$ are real, non-negative functions of $\lambda$ such that $m_0 = m$, $Z = 1$, and $\lambda_0 = 0$ for $\lambda = 0$; the space-time variable $x \in \RRR$, we use the $(-1, 1, 1, 1)$ metric; and the external source $J(x)$ is an \it arbitrary, \rm real scalar field, cf.~\cite{Itzykson} equations (8.38)-(8.40). So this QFT involves in fact only two real coefficients, mass $m$ and coupling constant $\lambda$, and a real field $Z^{1/2}\polje(x)$.

In the absence of interaction (i.e.~with $\Lagi \equiv 0$), the Euler-Lagrange equation of the QFT action functional (\ref{QFTaction})  reads
\begin{equation}
   \partial^\mu \partial_\mu \polje(x) - m^2 \polje(x) = - \izvor(x) \,.
   \label{QFTfreeELequat}
\end{equation}
The following two covariant\footnote{We will refer to a solution owing to a source as covariant if the Lorentz-transformed solution is owing to the Lorentz-transformed source} solutions to this equation are of interest to us in the case of a point source, i.e., when $\izvor(x) = \delta^4(x)$ (cf.~\cite{Itzykson} subsection~1-3-1):

(A)~The Feynman-Stueckelberg solution
\begin{equation}
   \polF(x) \equiv (2\pi)^{-4} \lim_{\epsilon\to 0} \int d^4k \, e^{ikx} \propFF(k) 
   \label{QFTFSsol}
\end{equation}
with
\begin{equation}
   \propFF(k) \equiv (k^2 + m^2 - i\epsilon)^{-1}, \qquad \epsilon \searrow 0 \,, 
   		\quad k \in \RRR \,, 
   \label{QFTprop}
\end{equation}
as $\polF(x)$ equals the QFT spin 0 free-field propagator $\propQ(x)$.

(B)~The retarded solution
\begin{equation}
   \poljeR(x) = (2\pi)^{-4} \lim_{\epsilon\to 0} \int d^4k \, e^{ikx} (\vec{k}^2 - (k_0 + i\epsilon)^2 + m^2 )^{-1} \,.
   \label{QFTRsol}
\end{equation}

These two solutions provide an example of the problems posed by QFTs: (A)~The momentum space, spin 0 Feynman propagator $\propFF(k)$ does not vanish fast enough as $|k^2| \to \infty$ to make convergent all the integrals in Feyman diagrams that are needed to compute the QFT perturbative $S$-matrix. (B)~The retarded solution (\ref{QFTRsol}) models only relativistic effects\footnotemark[1] that propagate away from a point source at speeds equal or less than the speed of light. So it displays no AFEs; and taking account of general sources, and/or of the interaction $\Lagi(\polje)$, is not likely to change that.

In view of Ockham's razor\footnote{Ockham's razor requires that ``Entities are not to be multiplied beyond necessity''. This basic, tacit principle of theoretical physics was invoked already by Galileo, cf. e.g. Encyclopaedia Britannica 2003.} these two problems pose the basic question whether there is a finite theory of quantum scattering such that: (i)~It can be defined non-perturbatively. (ii)~Its perturbative Green functions do not require regularization. (iii)~It can reproduce all the experimentally verifiable results of QFT. (iv)~It can model AFEs. (v)~It has conventional properties of theories in contemporary classical physics; in particular, it is not in conflict with Einstein's relativity postulates\footnotemark[3], even though relativistic causality may not always be obeyed.

The theory we will consider suggests, for the first time, that there is a positive answer in four-dimensional space-time to this basic question in theoretical physics. In particular, it suggests that there is a path-integral-based finite theory of quantum scattering with a Lagrangian that (a)~is local in space-time and invariant under the same transformations as the resulting $S$-matrix, in particular, under the Poincar\'e group of symmetry transformations, and (b)~its Euler-Lagrange equations have solutions that exhibit AFEs and are not in conflict with Einstein's relativity postulates.

We consider this theoretical problem because, inspired by the following two comments, we believe its solution to be relevant for the study of fundamental interactions.
\begin{itemize}
\item[(A)]~According to t'Hooft \cite{Hooft2}, ``History tells us that if we hit upon some obstacle, even if it looks like a pure formality or just a technical complication, it should be carefully scrutinized. Nature might be telling us something, and we should find out what it is''.
\item[(B)]~According to Dirac \cite{Dirac}, ``One can distinguish between two main procedures for a theoretical physicist. One of them is to work from the experimental basis \ldots The other procedure is to work from the mathematical basis. One examines and criticizes the existing theory. One tries to pin-point the faults in it and then tries to remove them. The difficulty here is to remove the faults without destroying the very great successes of the existing theory''.
\end{itemize}
And this is what we will try to do with regard to ultraviolet divergences and AFEs.

In subsection~\ref{framework} we specify a formal framework for non-perturbative finite alternative theories (FATs) to the considered scalar QFT specified by the action functional (\ref{QFTaction}). To construct a FAT Lagrangian, only the free part of the QFT Lagrangian is changed and defined using an additional independent four-vector variable; but it remains local in space-time and Lorentz-invariant. We take as a starting point the path-integral formalism with the Lehmann-Symanzik-Zimmermann (LSZ) reduction formula, see e.g.\cite{Sterman}. The resulting FAT Green functions are defined in the continuous four-dimensional space-time.

In subsection~\ref{altpropprop} we collate the properties of a FAT spin 0 free-field propagator that enable us to prove without using auxiliar particles that: (i)~the perturbative expansions of FAT Green functions are well defined, i.e.~without ultraviolet divergencies; and (ii)~the perturbative FAT $S$-matrix is unitary and equals the perturbative QFT $S$-matrix for certain value of a parameter in the FAT Lagrangian, on which it depends continuously.

In section~\ref{example} we construct such a FAT Lagrangian that we can (a)~reproduce the perturbative $S$-matrix of the considered scalar QFT, and (b)~model associated AFEs without either being in conflict with Einstein's relativity postulates, or introducing non-local equations of motion or backward-in-time effects (cf.~\cite{Eberh1}, subsection~2.3.2). 

In section~\ref{fizika} we have collated comments on physical content and implications of the considered FAT action functional. In particular, we comment on the relation between QFT and the FAT to it, on the special theory of relativity and the physics underlying quantum scattering, on physical content in addition to quantum scattering such as AFEs and the arrow of time, and on the particular particles suggested by Feynman \cite{Feynm} as the unifying concept for description and modeling of physical universe.

The results of this paper suggest that there is a solution to the above two basic problems posed by quantum scattering. In particular, they suggest that the physics underlying quantum scattering can respect Einstein's relativity postulates.

\section{FATs}
\label{theory}

Motivated by Pauli \cite{Villa} and Ockham's razor, we believe in the quest for such non-perturbative FATs to QFTs that are based on the path-integral formalism and have propagators that (i)~do not need to be regularized, and (ii)~can be regarded as such modifications of QFT propagators that could better reflect the \it underlying physics \rm than any of the regularized QFT propagators; for some comments on this belief see~\cite{Isham}.\footnote{In addition, there are two conventional, formal approaches to ultraviolet divergencies by changing the calculation of divergent Feynman diagrams (a)~by various regularizations, or (b)~by using dispersive techniques, see e.g.~Aste and Trautman \cite{Aste} and references therein.} So we propose

\subsection{A formal framework for FATs}
\label{framework}

As an example let us construct a FAT to the considered scalar QFT with the action functional (\ref{QFTaction}). To this end we modify this QFT as we put forward in~\cite{mi001, mi002}:

(a)~We introduce real scalar fields of $x \in \RRR$ with a continuous index $p \in \RRR$, say $\Polje(x; p)$. Under the Lorentz transformation $x \to \Lambda x + a$, these fields, defined on the eight-dimensional $\RRR \times \RRR$, transform as follows:
\begin{equation}
   \Polje(x; p) \to U(\Lambda, a) \Polje(x; p) \equiv \Polje(\Lambda x + a; \Lambda p) \,.
   \label{Lorentzscalar}
\end{equation}

(b)~We introduce a scalar weight $\utez(p^2) \in (-\infty, \infty)$, and the local weighted sum of fields $\Polje(x; p)$,
\begin{equation}
   \polj(x) = \int d^4p\, \utez(p^2) \Polje(x; p) \,.
   \label{tranpolje}
\end{equation}
So $\polj(x)$ is a real scalar field of $x$.

(c)~We replace the QFT free-field Lagrangian $\Lagf(\polje, \partial_\mu \polje)$ with a FAT free-field Lagrangian $\LagAf(\Polje, \partial_\mu \Polje)$ that is (i)~real; (ii)~local in space-time, in the sense that its value at any space-time point depends only on the values of $\Polje$ and $\partial_\mu \Polje$ at this point; (iii)~homogeneous of degree 2, i.e.,
\begin{equation}
    \LagAf(\alpha\Polje, \alpha\partial_\mu \Polje)  =  \alpha^2 \LagAf( \Polje, \partial_\mu \Polje) 
    \label{Afreeact}
\end{equation}
for all $\alpha \in (-\infty, \infty)$; and (iv)~Lorentz-invariant, in the sense that
\begin{equation}
   \LagAf(U(\Lambda, a)\Polje, \partial_\mu U(\Lambda, a)\Polje) = \LagAf(\Polje, \partial_\mu \Polje)\Big|_{x \to U(\Lambda, a)x} ;
   \label{Lagrcovariance}
\end{equation}
for an example see equation~(\ref{AfreeLagr}).

(d)~In the integrand in (\ref{QFTaction}) we replace the remaining fields $\polje(x)$ with the local sum $\polj(x)$ and so obtain the FAT action functional
\begin{equation}
   \akcijaA[\Polje; \izvor] =  \int \! d^4x \, \Bigl ( \LagAf(\Polje, \partial_\mu \Polje)  - \Lagi(\polj) + \polj \izvor \Bigr) \,,
  \label{Aakcija}
\end{equation}
corresponding to the QFT action functional $\akcija[\polje; \izvor]$. So the FAT Lagrangian in (\ref{Aakcija}) is local in space-time and defined by functions of eight independent variables $x$ and $p$; but in contrast with string theories, \it its interaction and source terms depend solely on the space-time variable $x$. \rm By (\ref{Lorentzscalar}), (\ref{tranpolje}) and (\ref{Lagrcovariance}), the FAT action functional (\ref{Aakcija}) is \it Lorentz-invariant \rm in the sense that
\begin{equation}
   \akcijaA[U(\Lambda,a)\Polje; U(\Lambda,a)\izvor] =  \akcijaA[\Polje; \izvor] \,.
   \label{actioninvariance}
\end{equation}

The scalar field $\izvor$ in (\ref{Aakcija}) does not depend on the index $p$. So on the analogy of QFT we can define the FAT, connected $n$-point Green functions $\GfA(x_1,\ldots, x_n)$ in the four-dimensional space-time $\RRR$ by functional derivatives of the generating functional
\begin{equation}
   Z_A[J] = \int \prod_{x,p} d\Polje(x,p) e^{ i \akcijaA [\Polje; \izvor] } \,;
  \label{Agenfunc}
\end{equation}
i.e.,
\begin{equation}
  \GfA(x_1,\ldots, x_n) = \left. { (-i)^{n} \delta^n \ln Z_A[J] \over\delta J(x_1) \cdots \delta J(x_n) }\, \right|_{J = 0} \,.
  \label{Greendef}
\end{equation}
By (\ref{Lorentzscalar})--(\ref{Agenfunc}), the FAT Green functions (\ref{Greendef}) are Lorentz-invariant in the sense that
\begin{equation}
   \GfA(U(\Lambda,a) x_1, \ldots, U(\Lambda,a) x_n ) = \GfA(x_1,\ldots, x_n) \,.
   \label{Greencovariance}
\end{equation}
And then we define the FAT $S$-matrix, $\SmA$, in terms of $\GfA$ by means of the LSZ reduction formula.

So, like the QFT nonperturbative path-integral formalism, also the corresponding FAT nonperturbative formalism (\ref{Aakcija})--(\ref{Greendef}) is \it local in space-time and Lorentz covariant. \rm Through such non-perturbative FAT formalism we generalized the perturbative method of unitary regulators invented by t'Hooft and Veltman in~\cite{Hooft} section~5.

\subsection{FAT spin 0 free-field propagator}
\label{altpropprop}

Let us denote by $\PolX(x;p,\izvor)$ a free field owing to the source $\izvor$, i.e., a scalar solution to the Euler-Lagrange equations of the FAT action functional $\akcijaA[\Polje;\izvor]$ with $\Lagi \equiv 0$.\footnote{Note that $\Lagi(\polj) \equiv 0$ for all $\polj$ only if $\lambda_0 = 0$, $Z = 1$, and $m_0 = m$, by (\ref{QFTintlagr}).} Changing the variables of functional integrals in (\ref{Agenfunc}) from $\Polje(x;p)$ to $\Polje(x;p) - \PolX(x;p,\izvor)$ and taking the properties of $\LagAf(\Polje, \partial_\mu\Polje)$ into account, we can write the generating functional (\ref{Agenfunc}) if $\Lagi \equiv 0$ as
\begin{equation}
   Z_A[J] = N_A \, \exp \left \{ {\textstyle{i\over2}} \int d^4x d^4y \, \izvor(x) \propO(x-y) \izvor(y) \right \}  \,,
   \label{Agenfreefunctrans}
\end{equation}
where $N_A$ is a $\izvor$-independent factor, and $\propO(x)$ is the local sum (\ref{tranpolje}) of a free field $\PolX(x;p,\delta^4(x))$, i.e.,
\begin{equation}
   \propO(x) \equiv \int d^4p \, \utez(p^2) \PolX(x;p,\delta^4(x)) \,.
   \label{polXdef}
\end{equation}

If a free field $\PolX(x;p,\delta^4(x))$ is such that $\propO(x)$ is a Lorentz-invariant scalar field of $x$ in the sense that
\begin{equation}
   \propO(\Lambda x) = \propO(x)  \,,
   \label{propminus}
\end{equation}
then $\propO(-x) = \propO(x)$, and by (\ref{Greendef}) and (\ref{Agenfreefunctrans}), in the absence of interaction, the FAT two-point Green function
\begin{equation}
   \GFAT(x,y) = -i\propO(x-y) \,.
   \label{GTFAfree}
\end{equation}
And there are diagrammatic perturbative expansions of FAT Green functions $\GfA$ that equal the diagrammatic perturbative expansions of QFT Green functions, with the QFT spin 0 free-field propagator $\propQ(x)$ replaced with $\propO(x)$, while vertices remain unchanged. 

By analogy with $\propQ(x)$, we will refer to $\propO(x)$ as a FAT spin 0 free-field propagator. We will denote it by $\propA(x)$, (A)~if there are no \it ultraviolet divergencies \rm in the perturbative expansions of $\GfA$, and (B)~if the perturbative FAT $S$-matrix (i)~involves solely identical spin 0 particles, (ii)~is \it unitary\rm, (iii)~depends continuously on a physical parameter, and (iv)~is \it equal \rm to the perturbative QFT $S$-matrix for a special value of this parameter.

We found out \cite{mi001,mi002} that to this end it seems to suffice that the Fourier transform $\propAF(k)$ of $\propA(x)$ equals the momentum space, spin 0 Feynman propagator $\propFF(k)$ multiplied by a suitable factor $\fakt(k^2 - i\epsilon)$, i.e.,
\begin{equation}
   \propAF(k) = (k^2 + m^2 - i\epsilon )^{-1} \fakt(k^2 - i\epsilon) \,, \qquad \epsilon \searrow 0 \,, \quad k \in \RRRR 
   \label{propagator}
\end{equation}
(which implies (\ref{propminus})), where the factor $\fakt(z)$ has the following analytical properties: (a)~it is an analytic function of $z \in \CC$ except somewhere along the segment $z \le z_d < -9m^2$ of the real axis; (b)~$\fakt(-m^2) = 1$; (c)~$\fakt(z)$ is real for all real $z > z_d$, so that $\fakt(z^*) = \fakt^*(z)$; (d)~we can estimate that for all $z \in \CC$, 
\begin{equation}
   | \fakt(z) | \le a_0 ( 1 + | z |^{3/2} )^{-1} \,, \qquad a_0 > 0 \,;
   \label{faktest}
\end{equation}
and that for any real $z_0 > z_d $ the derivatives $\fakt^{(n)}(z)$ of $\fakt(z)$ are such that
\begin{equation}
   \sup_{ z \in \CC, \Re z \ge z_0} (1 + |z|^{3/2}) (1 + |z|^n) | \fakt^{(n)}(z) | < \infty \,, \quad n = 1, 2, \ldots ;
   \label{ocenaodvoda}
\end{equation}
(e)~the coefficients of $\fakt(z)$ depend on a positive cut-off parameter $\Lambda$ so that for any $\Lambda \ge \Lambda_0$, $\Lambda_0 > 0$, the factor $\fakt(z)$ has properties (a)--(d) with the constants $z_d$ and $a_0$ independent of $\Lambda$, and we can estimate that in the asymptote of infinite cut-off parameter:
\begin{displaymath}
   \sup_{|z| < z_0} |\fakt^{(n)}(z) - \delta_{n0}| \to 0 \quad{\rm as} \quad \Lambda \to \infty \quad\hbox{for any}\quad z_0 > 0 , 
                     \quad n = 0,1,\ldots, 
\end{displaymath}
and
\begin{equation}
   \sup_{\Lambda > \Lambda_0} \sup_{z > 0} | z^n \fakt^{(n)} (z)| < \infty \qquad\hbox{for} \quad n = 0,1,\ldots .
   \label{regucond0}
\end{equation}

Relations (\ref{propagator}) and (\ref{faktest}), (\ref{propagator})--(\ref{ocenaodvoda}), and (\ref{propagator})--(\ref{regucond0}) are useful for proving the absence of ultraviolet divergencies and of auxiliary particles, perturbative unitarity, or approximability of the QFT perturbative S-matrix by the FAT one.

We will show in subsection~\ref{retcase} that there are such FAT free-field Lagrangians that their Euler-Lagrange equations have a solution that generates through (\ref{polXdef}) a FAT spin 0 free-field propagator with properties (\ref{propagator})--(\ref{regucond0}).

\subsection{Remarks}

\bf (a)~Comparison with other propagators. \rm Within the conventional QFT based on the canonical formalism, each complete, momentum-space, spin 0 Feynman propagator has the analytic properties (a)--(c) and (e) in subsection~\ref{altpropprop} but not (d); in particular, \it it has a cut, \rm as pointed out by Weinberg in the footnote in~\cite{Weinberg} p.~460. Whereas the Pauli-Villars regularizations result in propagators that have all the preceding properties except that the estimate (\ref{faktest}) is not valid for all $z \in \CC$. So as such, the assumed analytic properties (\ref{propagator})--(\ref{regucond0}) of the FAT propagator $\propAF(k)$ are not unusual.

\bf (b)~Unitarity of the FAT $S$-matrix. \rm According to Weinberg: ``Although the path-integral formalism provides us with manifestly Lorentz-invariant diagrammatic rules, it does not make clear why the S-matrix calculated this way is unitary. As far as I know, the only way to show that the path-integral formalism yields a unitary S-matrix is to use it to reconstruct the canonical formalism, in which unitarity is obvious.'', cf.~\cite{Weinberg} chapter~9. And that is the key problem with establishing conditions on the FAT Lagrangian for unitarity of the FAT $S$-matrix, since the canonical formalism can yield only a theory with ultraviolet divergencies, cf.~\cite{mi001} section~1. So in~\cite{mi001,mi002} we resorted to considering the unitarity of the perturbative FAT $S$-matrix, and searched for the properties of the FAT spin 0 free-field propagators that are sufficient to prove it; and we figured out the properties (a)--(e) given in subsection~\ref{altpropprop}. The question of properties that are both sufficient and necessary for proving the perturbative unitarity remains open and of general interest in view of the above Weinberg's remark about the known conditions on the Lagrangian for unitarity of the $S$-matrix in the path-integral formalism.

\section{A transport-theoretic example of a FAT and associated AFEs}
\label{example}

In what follows we show that one can construct such a FAT to the considered scalar QFT that there are free-field solutions to its Euler-Lagrange equations that exhibit AFEs. To this end, we use the following local, Lorentz-invariant FAT free-field Lagrangian
\begin{equation}
   \LagAf( \Polje, \partial_\mu \Polje) =  \half  \int \! d^4 p \, \Polje(x, -p) [ p^\mu \partial_\mu + \prep (p^2) ] \Polje(x, p)
                    - \half q \polj^2  \,,
   \label{AfreeLagr}
\end{equation}
where $q$ is a real coefficient; $\prep(y)$ is a real, bounded function of $y \in (-\infty, \infty)$, yet to be specified, such that $\utez^2(y)/ \prep(y)$ is finite at $y = 0$. Physical motivations for choosing this FAT free-field Lagrangian of the transport-theoretic kind are given in~\cite{mi003, mi004}. In subsections \ref{FSfree} and \ref{retcase} we will show there are such $\prep(y)$, $\utez(y)$ and $q$ that the action functional (\ref{Aakcija}) with (\ref{AfreeLagr}) determines a FAT and associated AFEs.

\subsection{Basic relations for FAT free fields}
\label{freeEL}

The FAT free fields $\PolX(x;p,\izvor)$, introduced in subsection~\ref{altpropprop}, are solutions to the Euler-Lagrange equations of the FAT action functional $\akcijaA[\Polje; \izvor]$ in the absence of interaction, i.e.~with $\Lagi \equiv 0$. So, for the FAT defined by (\ref{Aakcija}) and (\ref{AfreeLagr}), we have
\begin{equation}
   p^\mu \partial_\mu \PolX(x;p,\izvor) = q \utez(p^2) \polj(x;\izvor) - \prep(p^2) \PolX(x;p,\izvor) - \utez(p^2) \izvor(x) 
   \label{freeELequat}
\end{equation}
with
\begin{equation}
   \polj(x;\izvor) \equiv \int d^4p \, \utez(p^2) \PolX(x;p,\izvor) \,.
   \label{seenavsota}
\end{equation}
This Lorentz-invariant integro-differential equation is equivalent to the following integral relation between a free field $\PolX(x;p,\izvor)$, its local weighted sum $\polj(x;\izvor)$, and their source $\izvor(x)$:
\begin{eqnarray}
   \PolX(x;p,\izvor) &=& e^{-\prep(p^2) y_1} \PolX(x - y_1 p; p,\izvor) \nonumber \\
        &&+ \utez(p^2) \int_0^{y_1} \!dy\, e^{-\prep(p^2) y} [ q \polj(x - yp; \izvor) - \izvor(x-yp)] \,,
   \label{freeobrat}
\end{eqnarray}
which is valid for any $y_1 \in (-\infty, \infty)$, see, e.g.~\cite{Willi} subsection~2.3. 

Proceeding as in~\cite{mi004} we can compute from (\ref{freeELequat}) the following Lorentz-invariant relation between the space-time Fourier transforms $\poljF(k;\izvor)$ and $\izvorF(k)$ of the field $\polj(x;\izvor)$ and its source $\izvor(x)$:
\begin{equation}
   \poljF(k;\izvor) =  \Bigl( \int d^4p\, [ik\cdot p + \prep(p^2)]^{-1} [\utez(p^2)]^2 \Bigr) [ q\poljF(k;\izvor) - \izvorF(k) ] \,.
   \label{primerGreen}
\end{equation}
The space-time Fourier inverse of the solution $\poljF(k;\izvor)$ to (\ref{primerGreen}), and particular boundary and initial conditions for $\PolX(x;p,\izvor)$, determine by (\ref{freeobrat}) the specific FAT free fields $\PolX(x;p,\izvor)$.

\subsection{A particular  case}

To show by a simple example in subsections~\ref{FSfree} and \ref{retcase} that there are action functionals (\ref{Aakcija}) with (\ref{AfreeLagr}) that define a FAT and associated AFEs: (i)~Let the function
\begin{equation}
   \prep(y) = (-1)^j v_j \sqrt{y} \qquad\hbox{for}\qquad y \in (j-1,j] \,, \quad j= 1, 2, 3, 4, 
   \label{examplet}
\end{equation}
where $v_j = [ 1 + (2j - 5)\eta)] \Lambda $, with parameters $ \eta \in (0, 1/3)$ and $\Lambda \ge \Lambda_0 > m/(1 - 3\eta)$; and let $u(y) \ne 0$ for $y = 0$ and for $y > 4$. (ii)~Let
\begin{equation}
   q = { 2m^2 r_+ \over 2r_+ + m^2 r_-}  
   \label{primerkonst}
\end{equation}  
with
\begin{equation}
   r_\pm = \sum_{j=1}^4 d_j ( v_j^2 - m^2 )^{\pm 1/2} \,, \qquad
   d_j = (-1)^j (4 - |2j - 5|) \,. 
   \label{primerkonstante} 
\end{equation}  
(iii)~Let $\utez(y) = 0$ for $y > 4$, and let $\utez(y)$ be such that 
\begin{equation}
   2\pi^2 \int_{j-1}^j \utez^2(y) \sqrt{y} \, dy = - { |d_j | m^2 \over 2 q r_+ } \,, \quad j= 1, 2, 3, 4,
   \label{examplef}
\end{equation}
where the right-hand side is positive if $\eta \in (0, 1/3)$ and $\Lambda \ge \Lambda_0$. 

In such a case, we can check by inspection that, for $k^2 \ne -m^2$, the solution to (\ref{primerGreen})--(\ref{examplef}) is
\begin{equation}
   \poljF(k;\izvor) = (k^2 + m^2 )^{-1} \fakt_1 (k^2) \izvorF(k) \,, 
   \label{primerpropag}
\end{equation}
where
\begin{eqnarray}
   \fakt_1 (z) &\equiv& (z + m^2)\Big/ \Bigl[ q - z \Bigl( 2\pi^2 \int_0^\infty \utez^2(y) \prep(y) [ \sqrt{ 1 + z y /\prep^2(y)} - 1 ] dy
                         \Bigr)^{-1}  \Bigr]  \nonumber \\
   & = &   \Big [ \sum_{j=1}^4 d_j ( \sqrt{v_j^2 + z} + v_j )^{-1} \Big ] \Big/ \Big [ 2q \sum_{j=1}^4 d_j ( \sqrt{v_j^2 - m^2} 
                            + v_j )^{-1}        \nonumber \\
                 & &\qquad \times ( \sqrt{v_j^2 + z} + v_j )^{-1} ( \sqrt{v_j^2 + z} + \sqrt{v_j^2 - m^2}\,\, )^{-1} \Big ] \,.
   \label{tranpropexample}
\end{eqnarray}
This factor $\fakt_1(z)$ has the following properties:

(i)~$\fakt_1(z)$ defined by (\ref{tranpropexample}) is an analytic function of $z$ except along the segment $z \le z_d \equiv - \Lambda^2 (1 -3\eta)^2 < -m^2 $ of the negative real axis for each $\eta \in (1/3 - 0.047, 1/3)$; to infer this, we take into account that $\Re \sqrt z \ge 0$ and that $\Lambda/(\sqrt{v_j^2 - m^2} + v_j) \to  \infty$ as $\eta \nearrow 1/3$ only if $j = 1$. 

(ii)~$\fakt_1(z)$ is real for $z \ge - v_1^2$, $\fakt_1(-m^2) = 1$, $\sup_z (1 + |z|^{3/2}) |\fakt_1(z)| < \infty$, and by maximum modulus theorem relations (\ref{ocenaodvoda}) are satisfied.

(iv)~Relations (\ref{regucond0}) are satisfied if we choose $\Lambda$ as the cut-off parameter. 

So $(k^2 + m^2)^{-1}\fakt_1(k^2)$ is an analytic function of $k^2$ but for a first-order pole at $k^2 = - m^2$ and a cut at $k^2 \le z_d$. And we can obtain from space-time Fourier transforms of  (\ref{primerpropag}) the Feynman-Stueckelberg kind and retarded solutions to (\ref{primerGreen})--(\ref{examplef}) by replacing in rhs.(\ref{primerpropag}) $k^2$ with $k^2 - i\epsilon$ or $k_0^2$ with $(k_0 + i\epsilon)^2$, respectively, cf.~\cite{Itzykson} section~1-3-1.

\subsection{FAT Feynman-Stuckelberg free field}
\label{FSfree}

If we apply Feynman's prescription $k^2 \to k^2 - i\epsilon$ to (\ref{primerpropag}) with $\izvorF(k) = 1$, we obtain the space-time Fourier transform of the Feynman-Stueckelberg kind of solution to (\ref{primerGreen})--(\ref{examplef}),
\begin{equation}
   \polSFe(k;\delta^4(x)) =  f_1(k^2 - i\epsilon)/ (k^2 + m^2 - i\epsilon) \,,  
   \qquad \epsilon \searrow 0 \,.
   \label{FSprop}
\end{equation}

Using $\polSe(x;\delta^4(x))$ and (\ref{freeobrat}) with $y_1$ replaced by $\retfakt(p) y_1$, where $\retfakt(p) \equiv \hbox{sgn} (\prep(p^2) )$, and then limiting $y_1 \to \infty$ and $\izvor(x) \to \delta^4(x)$, we obtain the following expression for the corresponding solution to (\ref{freeELequat}),
\begin{eqnarray}
   \PolSe(x;p,\delta^4(x)) &=&  -\retfakt(p) \utez(p^2) \int_0^{\infty} \!dy\, e^{-|\prep(p^2)| y}  \label{freeobratFS}  \\
   &&\qquad [ q \polSe(x - \retfakt(p) yp;\delta^4(x)) + \delta^4(x- \retfakt(p) yp)] \,, \nonumber
\end{eqnarray}
which we regard as the FAT Feynman-Stueckelberg free field. Note that $\polSe(x;\delta^4(x))$ and $\PolSe(x;p,\delta^4(x))$ are covariant,${}^4$ but they are not retarded; and $\PolSe(-x;-p,\delta^4(x)) = \PolSe(x;p,\delta^4(x))$, by (\ref{seenavsota}) and (\ref{freeobratFS}).

$\PolSe(x;p,\delta^4(x))$ is a solution to (\ref{freeELequat}) which generates by (\ref{polXdef}), (\ref{seenavsota}), (\ref{FSprop}), (\ref{tranpropexample}), and (\ref{propagator})--(\ref{regucond0}) a FAT spin 0 free-field propagator $\propAe(x)$ such that (i)~its space-time Fourier transform
\begin{equation}
	 \propAFe(k) = f_1(k^2 - i\epsilon) \Big/ (k^2 + m^2 - i\epsilon) \,, \qquad \epsilon \searrow 0 \,,
	 \label{posebnitransform}
\end{equation}
and (ii)~it determines the corresponding FAT Feynman-Stueckelberg free field through (\ref{freeobratFS}) in analogy to relations (\ref{QFTFSsol})--(\ref{QFTprop}). So the action functional (\ref{Aakcija}) with the free-field Lagrangian (\ref{AfreeLagr}) provides a FAT to the considered scalar QFT.

\subsection{FAT retarded free fields and AFEs}
\label{retcase}

Theoretical considerations that take account of conservation laws and analysis of experiments of the EPR kind by the Bell inequalities imply that there is always an effect owing to one measurement apparatus, say A at $(ct_a, \vec{r}_a)$, on the results obtained by another measurement apparatus, say B at $(ct_b, \vec{r}_b)$, $\vec{r}_b \ne \vec{r}_a$, $t_b > t_a$, no matter how large is the ratio $|\vec{r}_b - \vec{r}_a|/c(t_b - t_a)$ \cite{Eberh1}. So we presume that such physical effects may be arbitrarily fast but not backward-in-time,\footnotemark[1] and will model these AFEs by a solution to the FAT free-field equation (\ref{freeELequat}).

To this end, we are interested in such solutions, $\PolR(x;p,\izvor)$, to the equation (\ref{freeELequat}) that the effects of the source $\izvor(x)$ on $\PolR(x;p,\izvor)$ are retarded. So when $\izvor(x) = 0$ for all $t \le t_0  \in (-\infty, \infty)$, then
\begin{equation}
   \PolR(x;p,\izvor) = 0 \qquad \hbox{for all}\qquad t \le t_0   \,.  \label{retsolut} 
\end{equation}
And if $\izvor'(x)$ and $\izvor''(x)$ are equal up to $t = t'$, then the corresponding retarded solutions $\PolR(x;p,\izvor')$ and $\PolR(x;p,\izvor'')$ are also equal up to $t = t'$, since the equation (\ref{freeELequat}) is linear.

The retarded free field $\PolR(x;p,\izvor)$ is not a covariant\footnotemark[4] solution to the Lorentz-invariant equation (\ref{freeELequat}) if it models AFEs, since otherwise it would exhibit also backward-in-time effects, cf.~\cite{Aharo}.

By (\ref{retsolut}), the local weighted sum $\polR(x; \izvor)$ of $\PolR(x;p,\izvor)$,
\begin{equation}
   \polR(x; \izvor) = \int \! d^4p \, \utez(p^2) \PolR(x;p, \izvor) \,,
   \label{retaverage}
\end{equation}
satisfies relation analogous to (\ref{retsolut}), and the effects of $\izvor(x)$ on $\polR(x; \izvor)$ are retarded. So it follows from (\ref{primerpropag}) that the relation between $\polR(x,\izvor)$ and $\izvor(x)$ is as follows:
\begin{equation}
   \polR(x;\izvor) = (2\pi)^{-4} \lim_{\epsilon\to 0} \int d^4k \, e^{ikx} {\fakt(k^2) \over k^2 + m^2} \Big |_{k_0 \to k_0 + i\epsilon}
                                  \izvorF(k) \,.
   \label{retprop}
\end{equation}
The local weighted sum $\polR(x;\izvor)$ being a covariant retarded solution to (\ref{primerGreen}), it describes relativistic effects.\footnotemark[1]

By (\ref{retsolut}) and (\ref{freeobrat}), if $\izvor(x) = 0$ for all $t \le t_0$, we can express for $p^0 \ne 0$ the retarded solution $\PolR(x;p, \izvor)$ to (\ref{freeELequat}) in terms of its source $\izvor(x)$ and the local weighted sum $\polR(x;\izvor)$ as follows:
\begin{equation}
   \PolR(x;p, \izvor) = \Theta(t - t_0) \utez(p^2) \int_0^{c(t-t_0)/p^0} \kern-25pt dy\, e^{- \prep(p^2) y} 
                   \{ \polR(x - y p; \izvor) + q \izvor(x - y p) \} 	
   \label{rw1sl}
\end{equation}
with $\Theta(t < 0) \equiv 0$ and $\Theta(t \ge 0) \equiv 1$. 

By (\ref{retprop})--(\ref{rw1sl}), the source $\izvor_a(x) \equiv \delta^{(4)}(x - (ct_a, \vec{r}_a))$: (i)~does not affect $\PolR(x;p, \izvor_a)$ if $t <t_a$, i.e., the responses of the system described by (\ref{freeELequat}), (\ref{examplet})--(\ref{examplef}) and (\ref{retsolut}) are retarded; and (ii)~affects $\PolR(x;p, \izvor_a)$ for some values of variable $p$ no matter how small is the time interval $t - t_a$ and/or how large is the distance $|\vec{r} - \vec{r}_a|$ if $t > t_a$, i.e., the physical system considered displays {\it everywhere arbitrary fast effects\/} owing to the source $\izvor_a(x)$. So $\PolR(x;p,\izvor_a)$ is not covariant, though the equation (\ref{freeELequat}) itself is Lorentz-invariant.

In the case of quantum scattering of identical spin 0 particles, which we model by the considered scalar FAT, there comes about a permanent information promptly available everywhere that a measuring apparatus has absorbed certain kind of particle. We propose to describe this infomation that such a particle has been absorbed at $x = x_a$ by $\PolR(x;p, \delta^4(x-x_a))$, motivated by its above properties. So we can model such quantum scattering and associated AFE by the same Lorentz-invariant and local action functional (\ref{Aakcija}) with (\ref{AfreeLagr}).

\section{Physical content and implications of FAT action functional}
\label{fizika}

\subsection{Relation between QFT and FAT} 
\label{zvezaQA}

A FAT to a given QFT is based on analogy with it on a generating functional. For the FAT interaction Lagrangian we prefer the QFT one, since its form is the main result of sixty years of research in quantum scattering. So we change only the QFT free-field Lagrangian.

When constructing the FAT free-field Lagrangians, we do not introduce formal auxiliary parameters that should be eventually disposed of when calculating the FAT $S$-matrix. In the asymptote of infinite cut-off parameter as $\Lambda \to \infty$, the perturbative FAT $S$-matrix equals the QFT one, which is the only QFT entity that directly relates to experimentally observable quantities. As $\Lambda \to \infty$, integrands in the perturbative FAT Green functions tend towards the QFT ones, most of which are ultraviolet divergent. Which suggests that in QFTs one resorts to regularizations of Green functions to somehow remedy the consequences of prematurely limiting some inherently physical cut-off parameter that is not used for modeling the available experimental results.

By analogy with alternatives $\propA(x)$ to the spin 0 Feynman propagator $\propQ(x)$, we have defined also alternatives to the spin $\half$ Feynman propagator and to the Feynman propagator in unitary gauge for massive spin 1 bosons, and provided an example of the corresponding FAT free-field Lagrangians \cite{mi005}: We used them to construct an example of a FAT to QED with massive photons in the unitary gauge. Similarly we can demonstrate that there are FATs to the standard model.

\subsection{Model of FAT and of AFEs and special relativity}
\label{FTESR}

The considered model of a FAT and of associated AFEs, based on the Lorentz-invariant action functional (\ref{Aakcija}) with (\ref{AfreeLagr}), is not in conflict with the Einstein relativity postulates as some regularizations in QFTs and models of AFEs are. So this model does not suggest that the physics underlying quantum scattering may be in conflict with these postulates. It shows also that locality and Lorentz-invariance of equations (\ref{freeELequat}) cannot preclude AFEs in the retarded solution $\PolR(x;p, \izvor)$, which exhibits certain interesting properties:
\begin{itemize}
\item[(i)]The law of evolution for $\PolR(x;p, \izvor)$ is given by the Euler-Lagrange equations (\ref{freeELequat}) and retardation condition (\ref{retsolut}), which are the same in all frames of reference. So this law is in accordance with Einstein's first relativity postulate.\footnotemark[3]
\item[(ii)]The values of $\PolR(x;p, \izvor)$ in one inertial frame of reference uniquely determine its values in any other one. However, $\PolR(x;p, \izvor)$ is not covariant, in contrast to its local sum $\polR(x; \izvor)$.
\end{itemize}
Some of these properties are required from a description of reality as specified by Eberhard~\cite{Eberh1} subsection~2.2.3.

\bf Causality and AFEs. \rm In the case of an experiment of the EPR kind with two measurement apparatuses A and B, considered in subsection~\ref{retcase}, we could conclude that A exerts effects on B because $t_b > t_a$. However, if $|\vec{r}_b - \vec{r}_a| > c(t_b - t_a)$, there are frames of reference where A is at $(ct_a', \vec{r}'_a)$ and B at such $(t_b', \vec{r}'_b)$ that $t_b' < t_a'$. So there, observing the same experiment we would conclude that B exerts effects on A. Thus in the case of faster-than-light effects causality is not a relativistic invariant as it is in the case of relativistic effects, cf. e.g. \cite{Kacser}.

\subsection{Physical content of FAT action functional in addition to quantum scattering}
\label{preko}

The FAT propagator $\propA(x)$ being the local sum (\ref{polXdef}) of a Feynman-Stuec\-kel\-berg solution $\PolS(x;p,\delta^4(x))$, the information about the FAT action functional (\ref{Aakcija}) that may be inferred from experimental results of quantum scattering is less complete than in the case of QFT. So the results of quantum scattering place much lesser restrictions on the FAT action functionals than on the QFT ones.\footnote{Which follows also from the fact that in contrast with QFTs, all FATs are without asymptotic completness since the K\"all\'en-Lehman spectral weight function of an alternative propagator is changing sign, see~\cite{mi001} equation~(22). And a LSZ $S$-matrix provides just the amplitudes for scattering between in and out particle states.} For instance, in our example (\ref{examplet})--(\ref{tranpropexample}) there are infinitely many functions $\utez(y)$ and $\prep(y)$ that result by (\ref{posebnitransform}) in the same FAT propagator $\propAe(x)$ that can reproduce the perturbative QFT $S$-matrix.

In contrast to the QFT action functional (\ref{QFTaction}), the considered transport-theoretic FAT action functional (\ref{Aakcija}) with (\ref{AfreeLagr}) provides a model of both quantum scattering and the associated AFEs. Furthermore, there is the arrow of time as the considered FAT action functional (\ref{Aakcija}) with (\ref{AfreeLagr}) is not invariant under the following time-reversal transformation:
\begin{equation}
   \Polje((ct,x),p) \to \Polje((-ct,x);p) \qquad\hbox{and}\qquad \izvor(ct,x) \to \izvor(-ct,x) \,.
   \label{slabTR}
\end{equation}

So there it is an extremely interesting open question what kind of physics ought a physically relevant FAT action functional contain in addition to that directly accessible by quantum scattering. FAT free-field Lagrangians that could take better account of the physics underlying quantum scattering than the QFT ones would be of interest e.g.: (i)~for avoiding anomalies, (ii)~for the study of non-perturbative phenomena (cf. \cite{Asorey}), (iii)~for evaluating the contribution of fundamental particles to the vacuum energy density (cf. \cite{Ossola}), (iv)~for determining an appropriate regularization of the one-loop effective action of QED (cf. \cite{Cho}), and (v)~for improving extraction of data from the QFT interaction Lagrangians. In this connection it is not of primary importance how convenient are the finite integrals in FAT Green functions for calculations---which is the main criterion for individually choosing the most convenient regularization for \it each \rm QFT Feynman integral, cf. e.g.~\cite{Weinberg} Ch.~11  and \cite{Deminov}.

\subsection{Feynman particles, X-ons, as the unifying concept for elementary interactions}
\label{Xons}

To our knowledge it was Feynman \cite{Feynm} who first suggested that the basic partial-differential equations of theoretical physics might be actually describing the macroscopic motion of some infinitesimal entities he called X-ons. In classical physics, such a motion is described by partial-differential equations of fluid dynamics, which can be extended to take some account of the underlying microscopic motion by the linearized Boltzmann integro-differential transport equation for a one-particle distribution. So we may regard the FAT free-field integro-differential equations (\ref{freeELequat})--(\ref{seenavsota}) as modeling in the linear approximation the transport of some infinitesimal entities, X-ons, with arbitrary four-momenta $p\in \RRR$, by the one-particle distribution $\PolR(x;p,\izvor)$. So that the macroscopic motion of X-ons is described in the linear approximation by $\polR(x;\izvor)$ and evolves almost according to the wave equation, by (\ref{retprop}) and (\ref{propagator})--(\ref{regucond0}).

In this connection, the following arguments of Polyakov \cite{Polya} are of interest: ``Elementary particles existing in nature resemble very much excitations of some complicated medium (ether). We do not know the detailed structure of the ether but we have learned a lot about effective Lagrangians for its low energy excitations. It is as if we knew nothing about the molecular structure of some liquid but did know the Navier-Stokes equation and could thus predict many exciting things. Clearly, there are lots of different possibilities at the molecular level leading to the same low energy picture.''

According to t'Hooft \cite{Hooft3}, ``We should not forget that quantum mechanics does not really describe what kind of dynamical phenomena are actually going on, but rather gives us probabilistic results. To me, it seems extremely plausible that \it any \rm reasonable theory for the dynamics at the Planck scale would lead to processes that are so complicated to describe, that one should expect \it apparently stochastic \rm fluctuations in any approximation theory describing the effects of all of this at much larger scales. It seems quite reasonable first to try a classical, deterministic theory for the Planck domain. One might speculate then that what we call quantum mechanics today, may be nothing else than an ingenious technique to handle this dynamics statistically.'' Now, kinetic theory is based on classical mechanics, with deterministic dynamical laws for particles (with arbitrary four-momenta with positive energies), and on probability theory, cf.~\cite{Willi}. Its results and t'Hooft's conjecture suggest that there is such a kinetic theory for Feynman's X-ons that in the limit of large degrees of freedom implies an adequate FAT of quantum scattering and also a model of associated AFEs. Such a theory would model reality as specified by Eberhard in \cite{Eberh1} subsection~2.2.3.

Constructing such a theory, we may note the points Einstein emphasized in a discussion about the significance of theory for observing physical phenomena. As related by Heisenberg \cite{Heise2} , (a)~``he insisted that it was the theory which decides about what can be observed'', and (b)~``Einstein had pointed out to me that it is really dangerous to say that one should only speak about observable quantities. Every reasonable theory will, besides all things which one can immediately observe, also give the possibility of observing other things more indirectly''.

\subsection*{Acknowledgement}

The authors appreciate many helpful comments and suggestions given by M. Polj\v sak and K.H.\ Rehren.

\vfill\eject

\end{document}